\documentclass{mem}
\usepackage{natbib}\usepackage{txfonts}\usepackage{balance}
\usepackage{graphicx}
\usepackage[a4paper,breaklinks,dvipdfm]{hyperref}
\idline{75}{282}
\begin{document}
\def\teff{$T\rm_{eff }$}
\def\kms{$\mathrm {km s}^{-1}$}

\title{
Doppler Tomography in Cataclysmic Variables: an historical perspective
}

   \subtitle{}

\author{
J. \,Echevarr\'\i{}a\inst{1}
          }

  \offprints{J. Echevarr\'\i{}a}

\institute{
Instituto de Astronom\'\i{}a, {\it Universidad Nacional Aut\'onoma de M\'exico},\\
Apartado Postal 70-264, M\'exico, D.F., M\'exico
\email{jer@astroscu.unam.mx}
}

\authorrunning{Echevarr\'\i{}a}

\titlerunning{Tomography in CV's}

\abstract{
To mark the half-century anniversary of this newly-born field of Cataclysmic Variables,
a special emphasis is made in this review, on the Doppler Effect as a tool in astrophysics.
The Doppler Effect was in fact, discovered almost 170 years ago, and has been since, one of the most
important tools which helped to develop modern astrophysics. We describe and discuss here, its use in Cataclysmic
Variables which, combined with another important tool, the tomography, first devised for medical purposes 70 years ago,
helped to devise the astronomical Doppler Tomography, developed only two decades ago. A discussion is made since the first 
trailed spectra provided a one dimensional analysis of these binaries; on the establishment of a 2D velocity profiling 
of the accretion discs; and unto modern techniques, which include Roche Tomography, time modulation and 3D imaging.

}

\maketitle{}

\section{Introduction}

\citet{doppler} published his monography {\it Über das farbige Licht der Doppelsterne und einiger anderer Gestirne des Himmels}
to explained the change in colour in binary stars and other stars in the sky. Although not yet based in spectral line shifting,
this triggered the works of Ballot in 1945 on sound waves, and that of Fizeau in 1948 on electromagnetic waves and the
change of the emitting frequency of an object, as it approaches or recedes. The Doppler effect is, without doubt of the the most important
tools in modern astrophysics. Particularly, in the study of spectroscopic binary stars, the Doppler Effect allows us to measure radial
velocity curves, and through them, calculate mass ratios and orbital periods, In the case of eclipsing binaries we are able to obtain
a full set of orbital parameters which include masses, ratios, orbital separation and inclination angle.

Another important scientific result is the medical tomograph, developed in the 1930s by the radiologist Vallebona,
who use the technique of projectional radiography. The mathematical basis for the reconstruction of a tomographic image was established by
\citet{radon}. The idea, as the name suggests, is to use several slice images taken around an object, and to reconstruct an image.
Doppler Tomography would therefore be a technique based on the reconstruction of an object based on doppler obtained, image slices, around an object.
Contrary to medical scans, in binary stars we take image slices, as the two stars revolve one orbital cycle around each other,
since we cannot go around them, like in the medical scans. It is obvious that such stars have to be short orbital periods systems, in order
to obtain results in a reasonable period of time. Interactive binaries are such a class of objects.

\section{Doppler Tomography of Cataclysmic Variables}

Although we are concerned here with Cataclysmic Variables (CV's), most of the results discussed in this review apply to
interactive binaries in general, i.e any class of double stars in which an exchange of matter occurs.
Doppler Tomography of Cataclysmic Variables should be understood as an image reconstruction of its components {\it via}
the doppler shift of the emission or absorption lines, visible in their spectra, along an orbital period. This {\it image}
is subject to interpretation, since we are only dealing with radial velocities and therefore we have to make
some assumptions about the location of the material in the stars and around them. Furthermore, as its is in most cases, we do not
know the inclination angle of the system and we observe only the projected velocity.

\subsection{Trailed spectra}

\begin{figure}[h!]
\resizebox{\hsize}{!}{\includegraphics[clip=true]{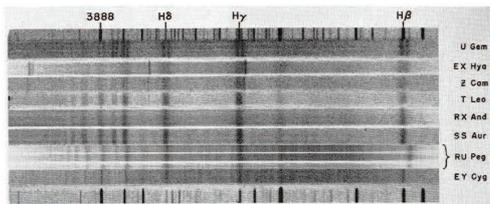}}
\caption{\footnotesize
Trailed spectra of Cataclysmic Variables
by \citet{kraft}.
}
\label{trail}
\end{figure}

The first tomograms of CV's are, in a sense, the early trailed spectra obtained
around the 1960s. An example of this is illustrated in Fig.~\ref{trail}.

%

The figure shows the photographic spectra of several Dwarf Novae in quiescence, obtained with
the prime-focus of the 200-in telescope.These {\it images} show, visually, the movement of
broad emission lines, the presence of asymmetric components and in some cases, like RU~Peg,
the presence of numerous absorption lines moving in anti-phase with the broad emission lines. The broad emission lines
give, in particular, a first picture of the accretion disc, its extension and degree of symmetry, while
the asymmetric or S-wave component, show the behavior of the hot spot in the disc, detected in photoelectric
light curves like in the case of U Gem \citep{kremin}. In the case of EY~Cyg, the system was thought to be near pole-on
and it took many years to detect the periodic radial velocity variation of its components \citet{eea07b}.

\subsection{1D Tomography}

\begin{figure}[h!]
\resizebox{\hsize}{!}{\includegraphics[clip=true]{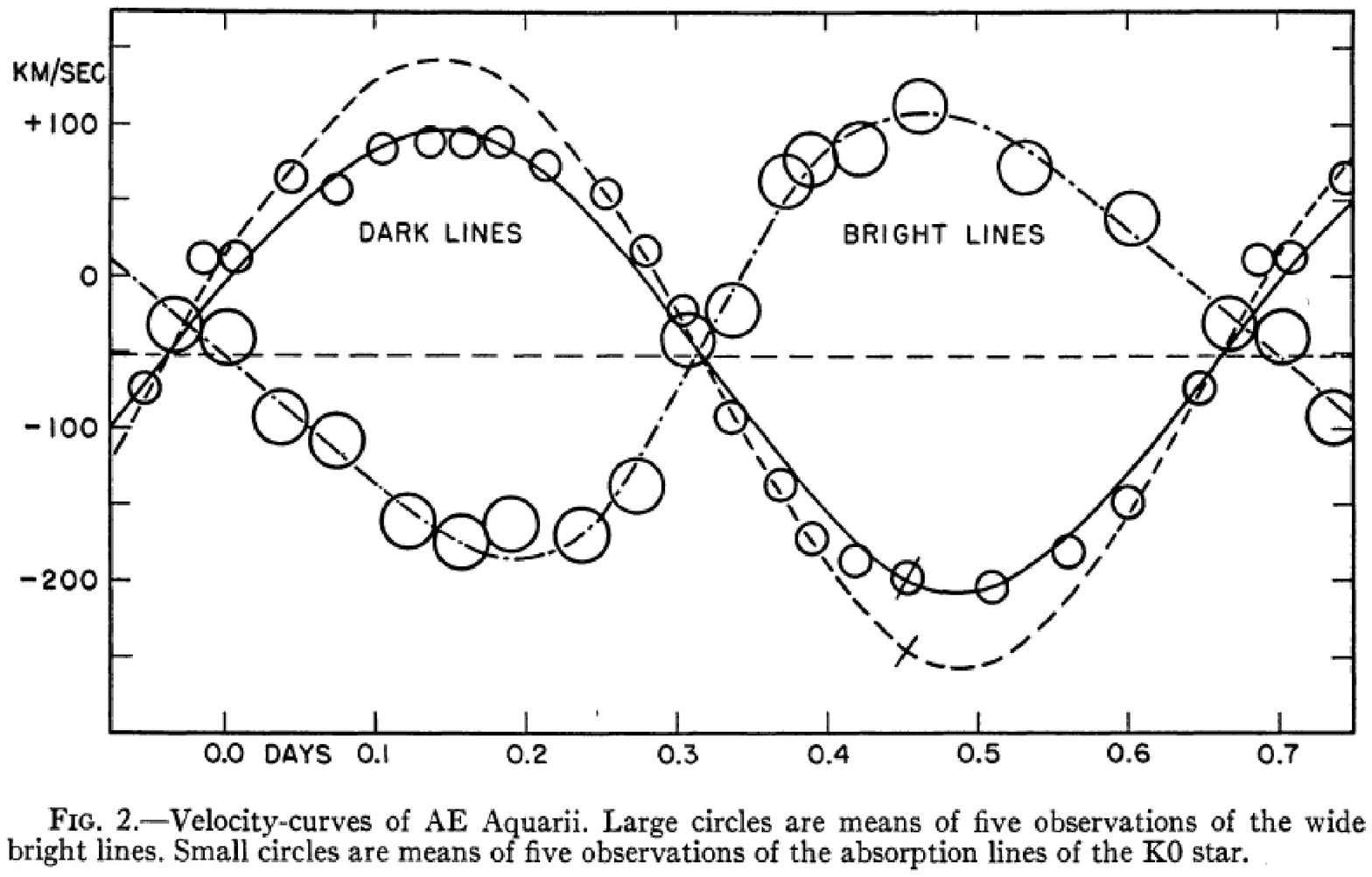}}
\caption{\footnotesize
 The radial velocity semi-amplitude of the components of AE~Aqr.
by \citet{joy}.
}
\label{joy}
\end{figure}

In a strict sense, any slice sections taken around an orbital period should be considered as a tomogram.
This is the case of the simplest analysis of the overall radial velocity measurements of the binary
components, which gives us a very basic, but fundamental picture of the system. This has been the basis
for obtaining orbital periods and mass ratios. One of the first radial velocity {\it picture} was
made by Joy in 1954 for AE~Aqr as shown in Fig.~\ref{joy}. The orbital period was overestimated by Joy,
but its value was later corrected by \citet{gap69}.

Radial velocity studies of CV's have been a keystone in many ways. They have provided the basics for
understanding their nature and variety \citet{warner}, as well as their formation and evolution \citet{ritter}.

\subsection{Classical Tomography}

In the late eighties \citet{mh88} proposed a method to reconstruct images of accretion discs, based
on the profiles of their emission lines. These are two-dimensional maps in velocity space. Their
proposed method, has been greatly successful. An extensive review can be found in \citet{marsh}. An Atlas of
Doppler Tomography for Cataclysmic Variables can be found in \citet{kea94}.

\begin{figure}[h!]
\centering
\includegraphics[trim=-1.5cm 0.25cm 0.0cm 0.0cm,clip,width=7cm]{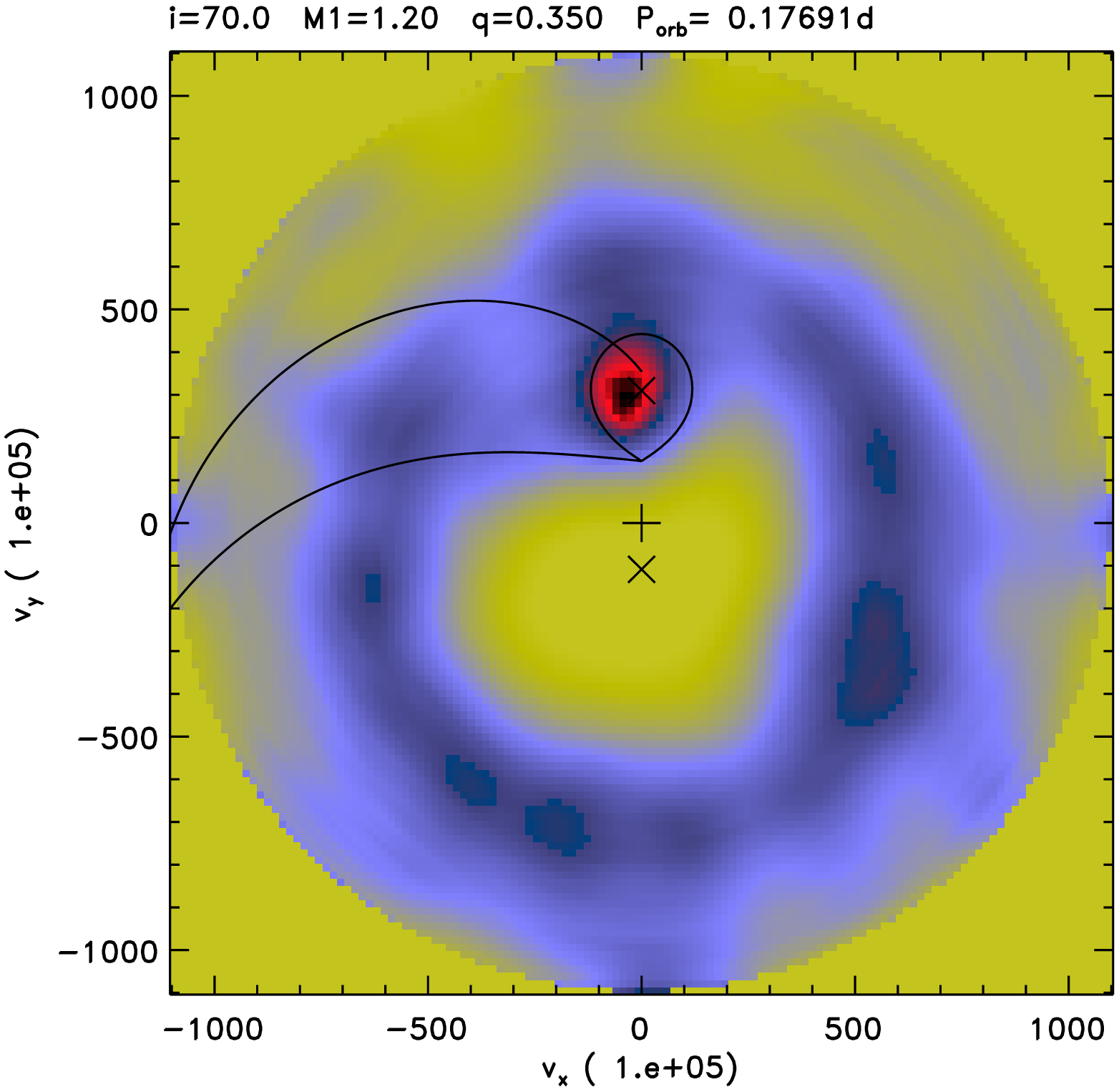}
\caption{\footnotesize
 An H${\alpha}$ velocity map of U Gem \citet{eea07a}).
}
\label{ugem}
\end{figure}

An example of a velocity space image is shown in Fig.~\ref{ugem}, for the  H${\alpha}$ emission line
in U~Geminorum \citet{eea07a}.

\begin{figure}[t!]
\centering
\includegraphics[trim=-1.5cm 0.25cm 0.0cm 0.0cm,clip,width=7cm]{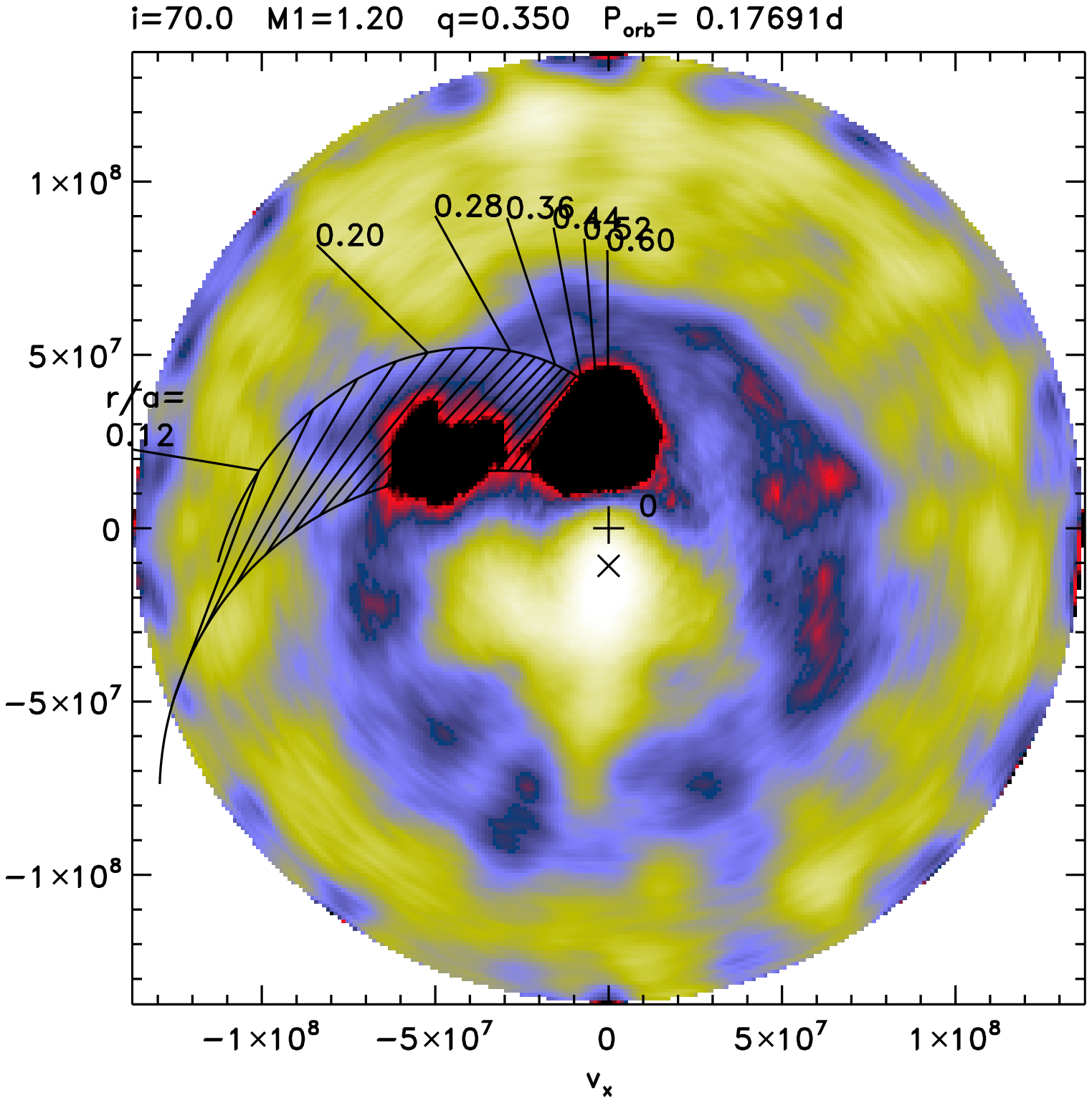}
\caption{\footnotesize
 Velocity map of U~Geminorum in 2006 \citet{eea12b}.
}
\label{ugem2}
\end{figure}

This tomogram has the typical layout for a CV. The Roche Lobe of the secondary
is shown, as well as the keplerian and stream trajectories of the transferred matter. It is important to stress
out that in this representation in space velocity, the inner disc corresponds to the outer geometrical boundary,
while the external disc, at the higher velocities has a cut-off at the boundary layer. It is therefore necessary
to interpret this map, as there is not a one to one correspondence between velocity and space. As an example of this
ambivalence we point out at the position of the hot spot in the diagram. Although the material seems to be located
in the forward face of the secondary, it is also possible that the emission comes from the $L_1$ point and has rapidly
acquired a keplerian velocity. This is unusual in U Gem, as we should have expected to see the hot spot much
further away, along the keplerian line, colliding with the disk to the left of the $V_y$ axis. However, at least in the
case of the Balmer lines, the hot-spot appears to come, at times, from a mixture of disc and stream velocities
\citet{mea90a}, often with an average velocity between them. This is clearly the case, shown in Fig.~\ref{ugem2}, from a 
tomogram derived for U~Gem \citep{eea12b}.

\subsubsection{On the Basic assumptions}

We also assume that all the visible material is in the orbital plane, which might not
be the real case, especially in polars, where the material is funneled into the white dwarf by the magnetic poles,
outside the orbital plane.  The principles of standard Doppler Tomography have been discussed by \citep{marsh2}. Some of them
are obviously contravened, like polar flow, emission from the mass donor and systems in outbursts.
Therefore, one has to be careful in the spatial interpretation of any tomogram.

\subsubsection{Achievements}

\begin{figure}[h!]
\centering
\includegraphics[trim=0.5cm 0.25cm 0.0cm 0.0cm,clip,width=6cm]{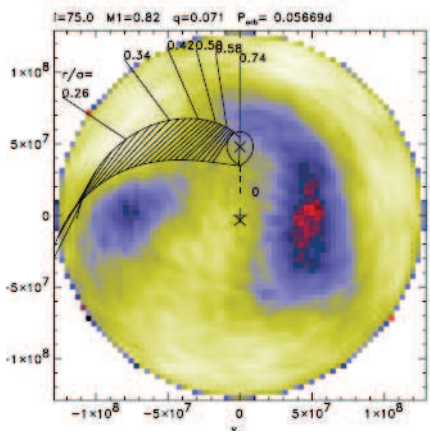}
\caption{\footnotesize
 He II tomogram of WZ~Sge during the 2001 super-outburst by \citet{bea02}.
}
\label{wz-baba}
\end{figure}

\begin{figure}[h!]
\centering
\includegraphics[angle=90,trim=0.0cm 0.0cm 0.0cm -2.0cm,clip,width=6cm]{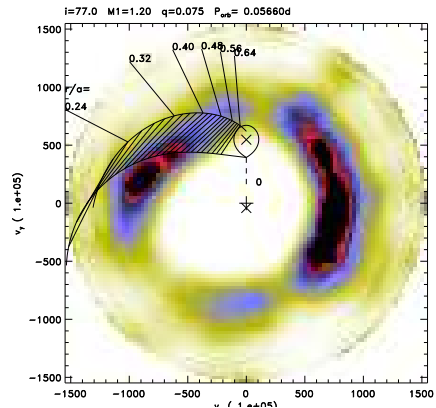}
\caption{\footnotesize
 $H\beta$ tomogram of WZ~Sge during super-outburst taken on 2001, July 6  \citep{eea12a}.
}
\label{wz-eche}
\end{figure}

\begin{figure}[h!]
\centering
\includegraphics[angle=90,trim=0.0cm 0.0cm 0.0cm -2.0cm,clip,width=6cm]{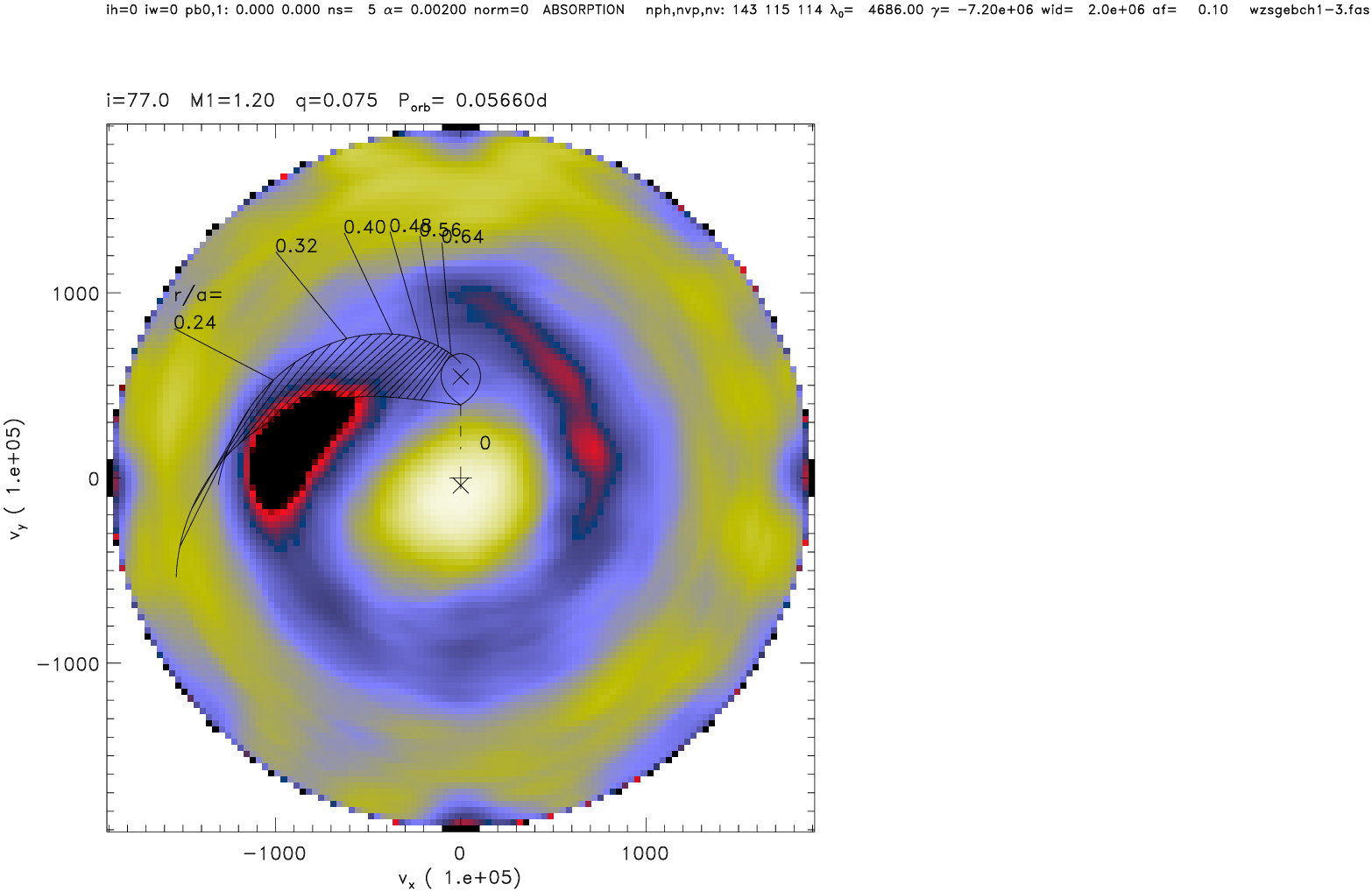}
\caption{\footnotesize
HeII tomogram of WZ~Sge during super-outburst taken on 2001, July 1-3  \citep{eea12a}.
}
\label{wz-heii}
\end{figure}

\begin{figure}[h!]
\centering
\includegraphics[trim=-1.5cm 0.25cm 0.0cm 0.0cm,clip,width=7cm]{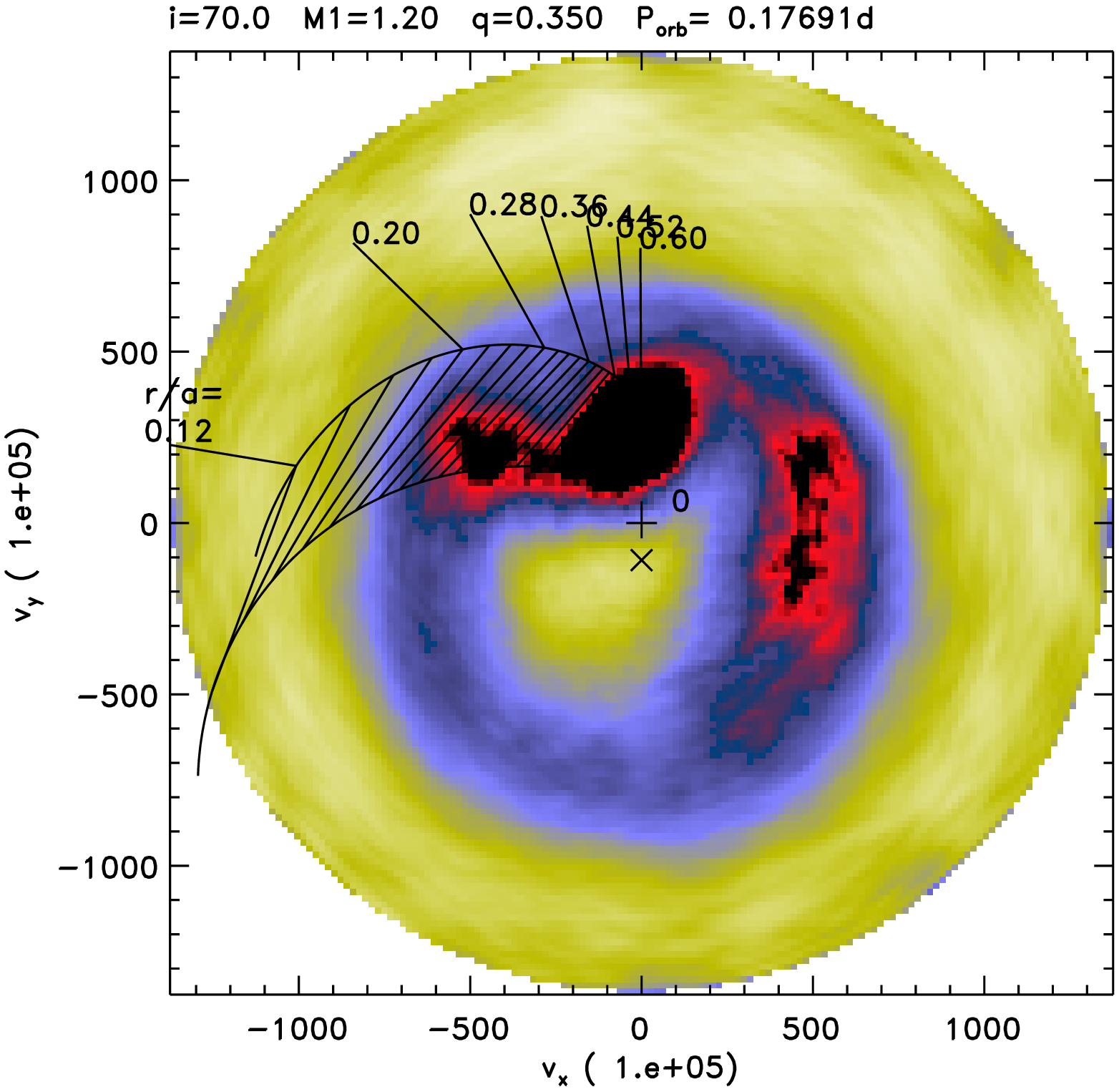}
\caption{\footnotesize
 Velocity map of U~Geminorum in 2008 \citet{eea12b}.
}
\label{ugem3}
\end{figure}

New discoveries have been achieved with Doppler Tomography, which include the presence of spiral structures in discs; bright-spots
from the secondaries stars; missing discs in nova-like systems; as well as stream emission in polars. The Dwarf Nova IP Peg, was the first system
found to have a spiral-arm structure during their outbursts \citep{sea97}. Since this important discovery, this spiral structure has been detected
in several systems. Among them are the nova-like V3885~Sgr \citep{hea05} and UX~UMa \citep{nea11}; WZ~Sge during its 2001 super-outburst
\citep{bea02} (Fig.~\ref{wz-baba}) and \citep{eea12a} (Figs.~\ref{wz-eche} and \ref{wz-heii}); and the dwarf nova U~Gem during outburst \citep{gr01}.
During quiescence \citet{ncv04} have found indications of a spiral structure in U~Gem. \citet{eea12b}
have also found that, at this stage, U~Gem has a complex behavior, including spiral structures as shown also in Fig.~\ref{ugem3}.


During the 2001 super-outburst of WZ~Sge, evidence has been found by \citet{pea02} and
\citet{eea12a} (see also Fig.~\ref{wz-heii}) that the hot-spot increases substantially. As remarked by Patterson,
it is possible that enhanced mass transfer from the secondary plays a major role in the eruption.

Bright-spots from the secondary stars have been detected in several systems like U~Gem (see \citet{mea90a}; \citet{uea06}; \citet{eea07a});
BY~Cam \citep{sea05}, and OY~Car in outburst \citep{hm96}, among others (see also Harlaftis and Marsh and references therein).
These emissions can be interpreted in some cases as mass transfer seeing through the $L_1$, which collide with a large accretion disc,
producing a hot-spot near the lagrangian point \citep{eea07a}; or from irradiation on the secondary star by the inner disc \citep{marsh}.
As pointed out by Harlaftis and Marsh, the origin of this emission, coming from all type of stars and in different states, has yet to be determined.

\begin{figure}[h!]
\centering
\includegraphics[trim=-1.5cm 0.25cm 0.0cm 0.0cm,clip,width=7cm]{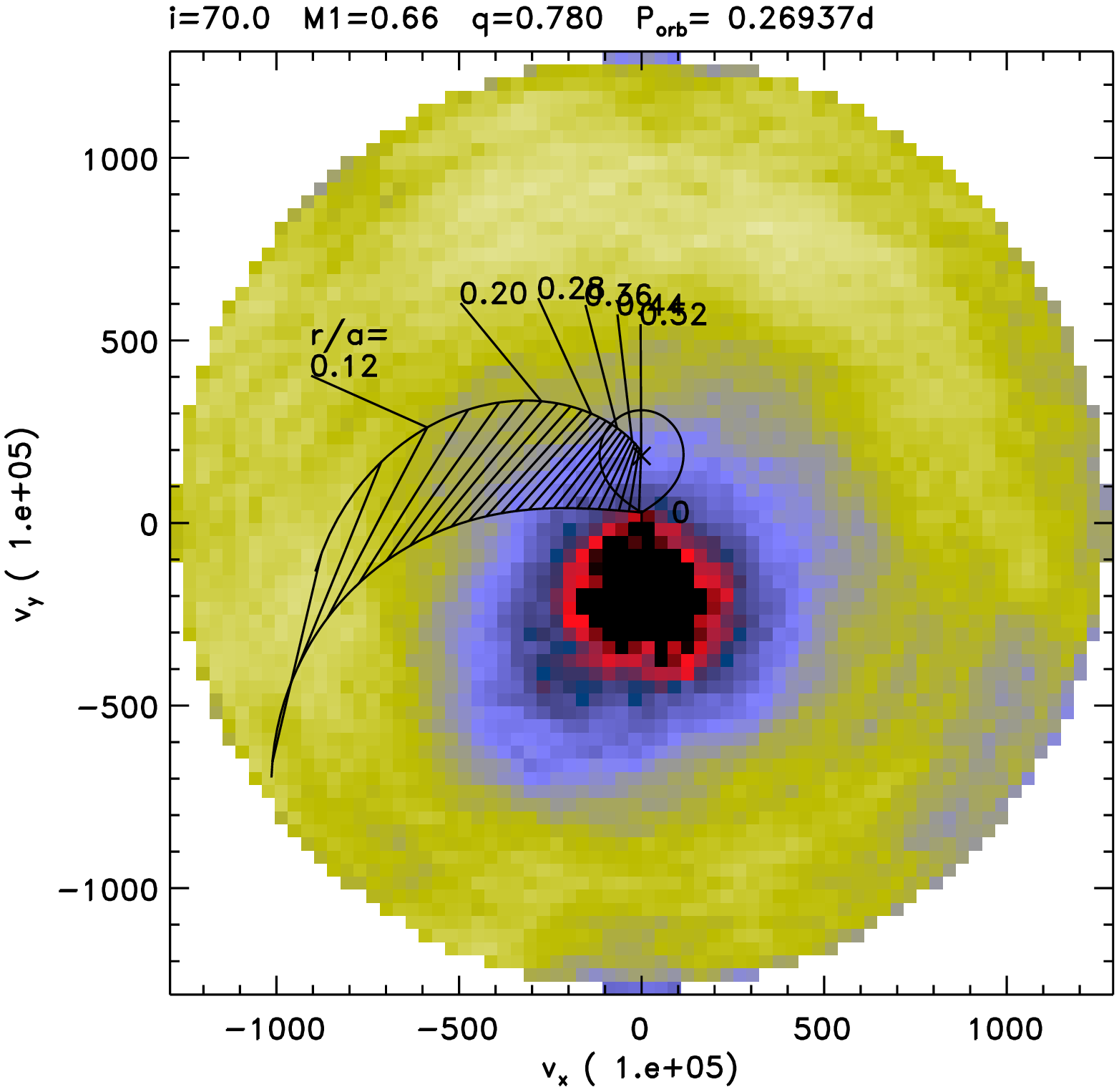}
\caption{\footnotesize
 HeII velocity map of J0644 \citet{hea12}.
}
\label{j0644}
\end{figure}

In some nova-like stars there is little evidence of a full accretion disc, like in UU~Aqr, V413~Aql, V363~Aur, AC~Cnc, VZ~Scl,
LX~Ser, SW~Sex, RW~Tri and SW~UMa (see \citet{kea94}). In some cases a ring is seen, but in many cases there is a large blob
of material only in the second and third quadrant of the disc or a dense blob near the location of the white dwarf. The high
transfer rates in nova-like systems is thought to be a probable cause of this, as the material is jettisoned along the stream
trajectory and returns to the disc at the opposite direction of the secondary star, producing either the large blob, or
captured in the vicinity of the white dwarf.
Such is the case of J0644+3344, a newly discovered eclipsing binary, classified as a CV by \citet{sea07} and discussed in
these proceedings by \citet{he12}. The $H\alpha$ and $H\beta$ maps show a large blob around the third quadrant. There is also
evidence of material near the white dwarf in the $H\alpha$ tomogram and, as in other objects, HeII shows
a strong and narrow emission near the vicinity of the white dwarf, as shown in Fig.~\ref{j0644} \citep{hea12}.

The characteristics of the system indicates that this is a SW~Sex type system rather than a UX~UMa one. Both classifications 
were suggested by \citet{sea07}.

\begin{figure}[t!]
\centering
\includegraphics[trim=0.0cm 0.25cm 0.0cm 0.0cm,clip,width=5cm]{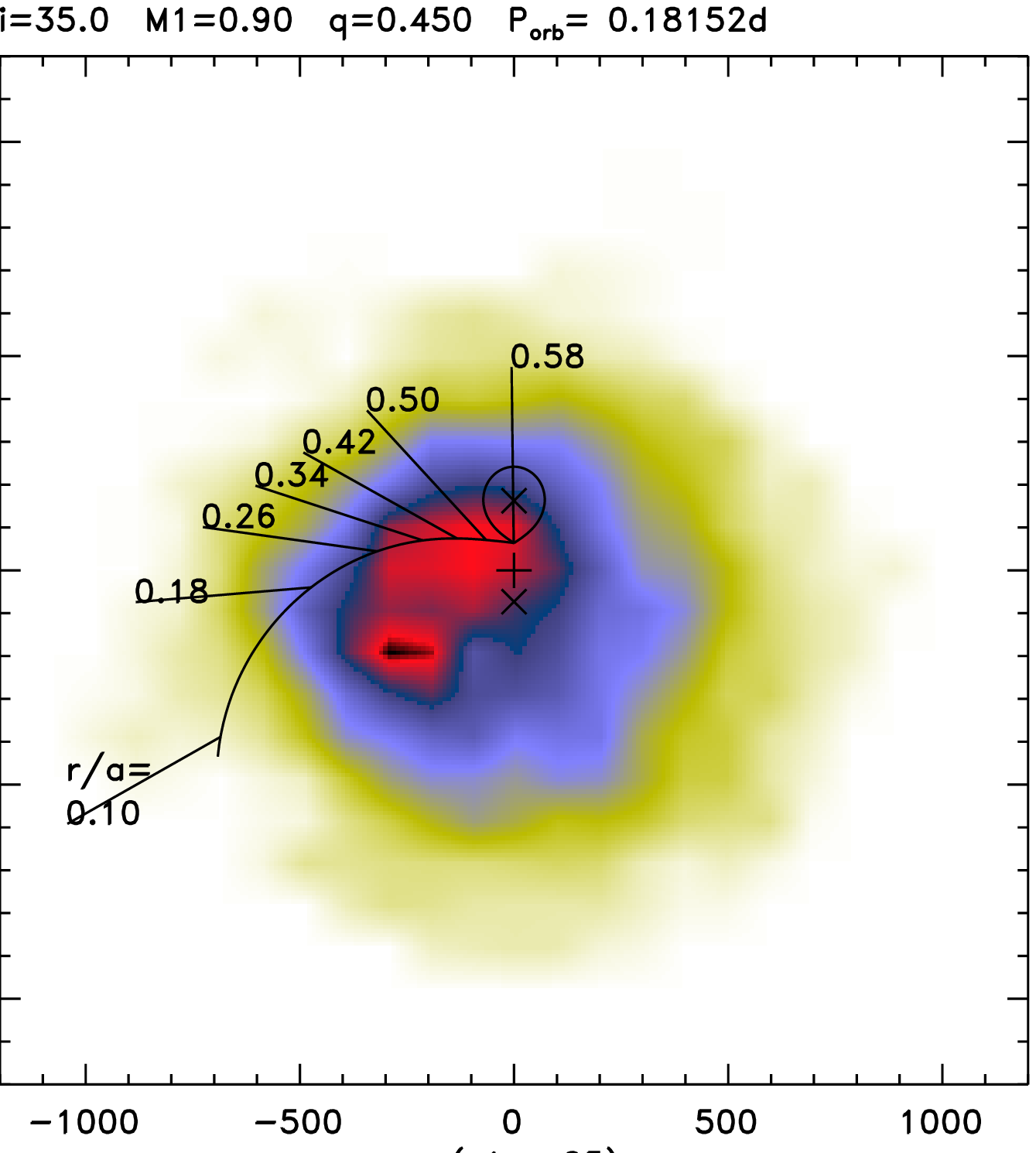}
\caption{\footnotesize
 $H\beta$ velocity map of V2306~Cyg \citet{zea01}.
}
\label{v2306-cyg}
\end{figure}

Doppler tomograms of polars have shown unmistakably the presence of stream-like flow unto the magnetic poles
of the primary stars, like in HU~Aqr \citep{hea99} and UZ~For \citep{sea99}. In the case of the asynchronous polar
BY Cam \citep{sea05} a curtain emission is rather detected, while in the semi-polar V2306~Cyg \citep{zea01} an accretion
ring is observed, as well as hot-spots caused by the X-ray beam and from the interaction of the mass transfer stream, as shown in
Fig.~\ref{v2306-cyg}.

We point out again, that what we {\it see} in these tomograms is an interpretation of where the observed material
is in the geometrical space of the binary. As discussed, one of the assumptions to construct classical Doppler
Tomography is to assume that all the material is in the orbital plane, which evidently in polars, might not be the case at all!

\subsection{Roche Tomography}

A tomographic application to study the secondary stars in Cataclysmic Variable has been developed in the last two decades.
Roche Tomography (\citet{rd94}; \citet{sm95}) use the absorption lines profile which is interactively fitted to a grid of
square specific fluxes lying on the surface the critical potential surface that defines the Roche Lobe. A full extent of the
procedure and basic assumptions can be found in \citet{dw01}. Roche Tomography from single line data like the Na I doublet, 
used in AM~Her, IP~Peg and QQ~Vul \citep{wea03}, reveal only partial information such as irradiation \citet{sm12}. 
Similar results have been obtained  in HU~Aqr by these authors, using the He II~4686 emission line, and on EX~Hya by \citet{br08}
using the NaI doublet in absorption and emission, as well as CaII~8498 in emission.

\begin{figure}[h!]
\centering
\includegraphics[trim=0.0cm 0.0cm 0.0cm 0.0cm,clip,width=6cm]{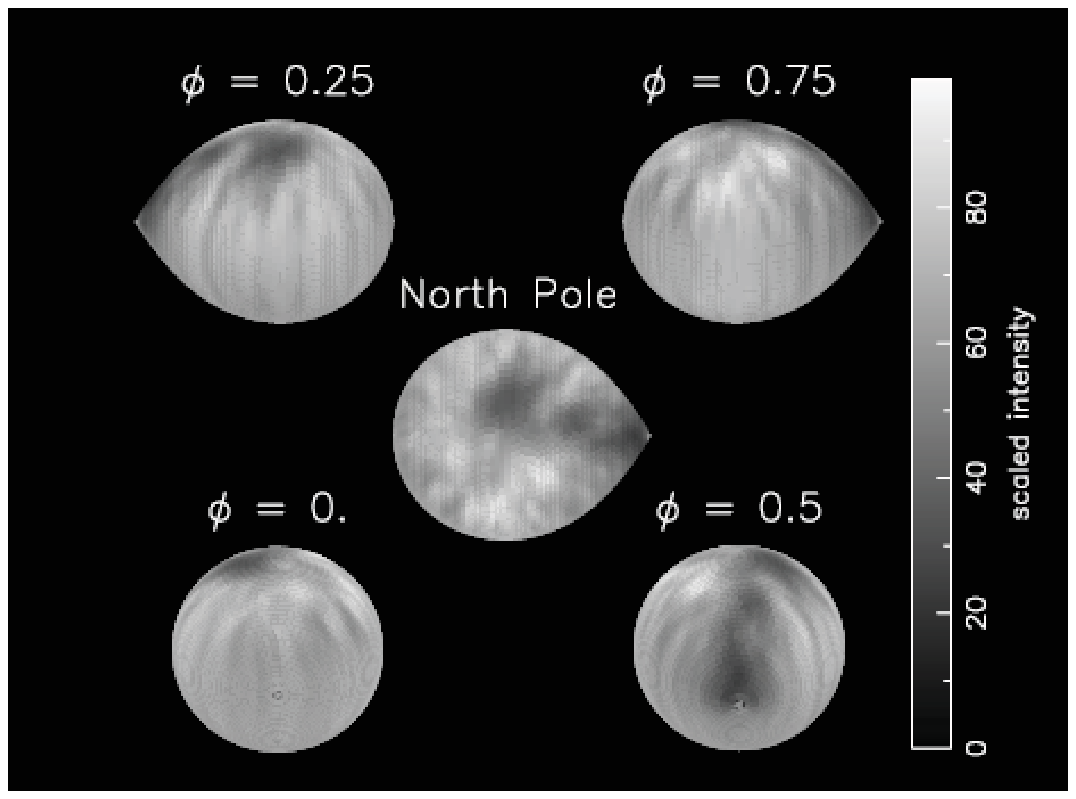}
\caption{\footnotesize
 Roche Tomography of AE~Aqr from \citet{wea06}.
}
\label{aeaqr}
\end{figure}

To detect spots on the secondary surface many spectral lines are needed \citep{dw01}, using techniques such as the Least-Squares 
Deconvolution (LSD), developed for single rapidly rotating active stars \citep{dea97}. First results have been obtained on AE~Aqr
by \citet{wea06} (see Fig.~\ref{aeaqr}). The secondary star shows several large, cool star-spots and the presence of a large, 
high-latitude spot, similar to that seen in rapidly rotating isolated stars. Similar results have been obtained on BV~Cen by 
\citet{wea07} and on RU~Peg by Dunford et al. 2001 (see \citet{sm12}).

\begin{figure}[h!]
\centering
\includegraphics[trim=0.0cm 0.0cm 0.0cm 0.0cm,clip,width=4cm]{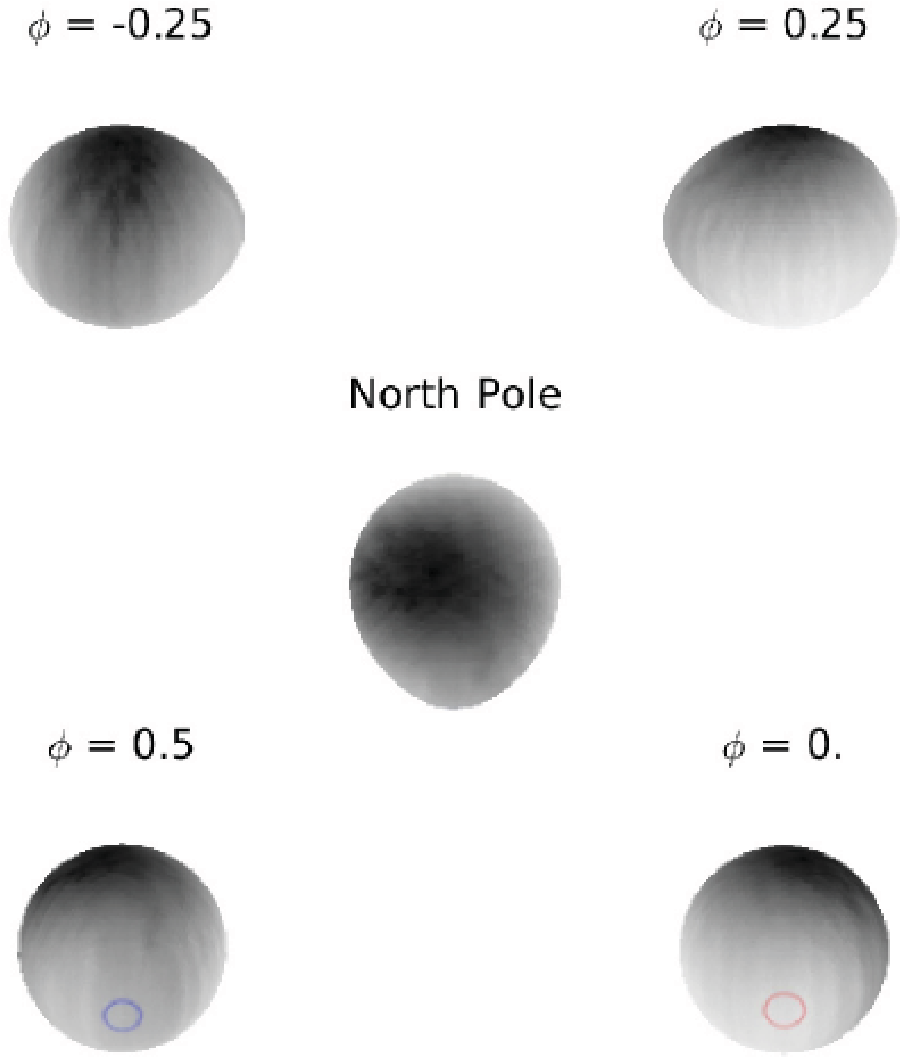}
\caption{\footnotesize
 Roche Tomography of LTT~560 from \citet{tea11}.
}
\label{ltt560}
\end{figure}
Although the Roche Tomography in Cataclysmic Variables is a powerful tool to study the secondary stars, it has the disadvantage,
compared to isolated or detached binary stars, that the accretion disc masks the signature of the absorption lines and therefore,
only large orbital period systems can be observed. For example, based on the Roche Tomography method by \citet{rd94}
and the LSD analysis, \citet{tea11} have found, in the detached post-common envelope binary LTT~560, that a large area of the secondary 
star on its leading side is covered by star spots, as shown in Fig.~\ref{ltt560}.

\subsection{Modulated and 3-Dimensional Tomography}

\begin{figure}[h!]
\centering
\includegraphics[trim=0.0cm 0.0cm 0.0cm 0.0cm,clip,width=5.5cm]{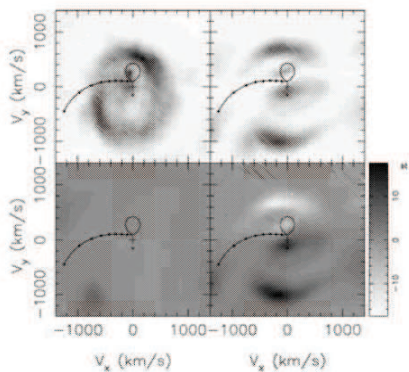}
\caption{\footnotesize
 Modulated Tomography of IP~Peg from \citet{st03}.
}
\label{ippeg-mod}
\end{figure}

New applications of Doppler Tomography have been developed. Among them, time modulated as well as 
three-dimensional (3D) Doppler Tomography. \citet{st03} has extended the classical tomography to a modulated 
emission line tomography by relaxing the requirement that the material is visible at all times and allowing
the mapping of time-dependent emission sources. He has applied a new code to IP~Peg during outburst and finds
(see Fig.~\ref{ippeg-mod}) that the asymmetric two-arm disc modulates strongly in terms of its sine and cosine amplitudes.

\citet{aea06} have also developed a three-dimensional (3D) tomography by relaxing, this time, the requirement that all the
material is all in the orbital plane, thus exploring the z-axial dependance. They find a high velocity stream across
the orbital plane for the Algol-type system U~CrB which has a well know high inclination angle. \citet{rea10} have also 
found evidence of out of the plane flow for the system RS~Vul.

\section{Conclusions}

We have reviewed, Doppler Tomography in Cataclysmic Variables, making a special emphasis on the
Doppler Effect as a tool in general in astrophysics, a discovery made more that a century and a half
years ago. Combined with another important tool, the tomography, devised for medical purposes during de middle of the last century,
Doppler Tomography has been devised for Cataclysmic Variables recently. A discussion was made, starting from the first trailed spectra
to the establishment of a 2D velocity profiling of the accretion discs and unto other techniques, which involve Roche Tomography of
the secondary stars, time modulated and 3D imaging tomography of the discs.

\section{Questions}

Question: Dmitry Kononov

As you state, I understand that the bright spot increase during the outburst
due to the increase mass transfer rate. But don't you suppose that it may happen
due to the redistribution of the intensity when the almost destroyed disc becomes
fainter and most of the contribution to the tomogram is from the shocks.

Answer: Echevarr\'\i{}a

It is the case in WZ~Sge that a well formed disc still remains near the peak of the outburst, nearly 10 days after maximum
as shown in Fig.~\ref{wz-heii}. This tomogram was constructed from spectra taken on 2001, 1-3 of July
and three days later, on July 6, the system still shows a spiral structure as shown in Fig.~\ref{wz-eche}.

Question: Dmitry Bisikalo

Could you comment on the perspectives of 3D tomography?

Answer: Echevarr\'\i{}a

The method has proved to be effective for very well know high inclinations systems. It has yet to be seen
if it can be applied other systems as well. The adjustment of a stretching effect, specifically a general
deconvolution of the image along the z-direction, presents a significant constraint on the full reconstruction of the 3D image.

\begin{acknowledgements}
Part of this work has been done with observations supported by the grant from DGAPA IN122409.
\end{acknowledgements}

\bibliographystyle{aa}

\end{document}